\documentclass[prl,twocolumn,superscriptaddress,showpacs]{revtex4-1}
\usepackage{amsmath,amssymb,graphicx,color,bm,soul}

\begin{document}

\author{T. C. H. Liew}
\affiliation{Division of Physics and Applied Physics, Nanyang Technological University
637371, Singapore}
\author{Y. G. Rubo}
\affiliation{Instituto de Energ\'{\i}as Renovables, Universidad Nacional Aut\'onoma de
M\'exico, Temixco, Morelos, 62580, Mexico}
\author{A. V. Kavokin}
\affiliation{Russian Quantum Center, 100, Novaya Str., Skolkovo, Moscow Region, 143025,
Russia.}
\affiliation{Physics and Astronomy School, University of Southampton, Highfield,
Southampton, SO171BJ, UK}
\title{Exciton-Polariton Oscillations in Real Space}
\date{\today}

\begin{abstract}
We introduce and model spin-Rabi oscillations based on exciton-polaritons in
semiconductor microcavities. The phase and polarization of oscillations can be controlled
by resonant coherent pulses and the propagation of oscillating domains gives rise to
phase-dependent interference patterns in real space. We show that interbranch
polariton-polariton scattering controls the propagation of oscillating domains, which can
be used to realize logic gates based on an analogue variable phase.
\end{abstract}



\maketitle

\textit{Introduction.---} Rabi oscillations are well-known for their role in
nuclear magnetic resonance devices and underly proposals for quantum
computing~\cite{Zoller1995,Leibrandt2009}. Following the archetypical
example of coherent and reversible energy transfer between atoms and light
in electromagnetic cavities~\cite{Kaluzny1983}, Rabi oscillations have been
achieved at the quantum level in a variety of systems, including: Josephson
junctions~\cite{Martinis2002,Yu2002}; electron spins in quantum dots~\cite%
{Petta2005,Koppens2006}; nuclear spin systems~\cite{Jelezko2004}; and
molecular magnets~\cite{Yang2012}.

Rabi oscillations were also observed in planar semiconductor systems, such
as quantum wells containing excitons~\cite{Schulzgen1998}, and semiconductor
microcavities containing exciton-polaritons~\cite{Norris1994}. Conversion of
spin-polarized excitons into circularly polarized photons and vice versa in
microcavities results in magnetization oscillations with terahertz
frequencies \cite{Brunetti2006}. Planar microcavities also allow the
ballistic transport of energy in space~\cite{High2013}, with
exciton-polaritons covering distances on the order of hundreds of microns~%
\cite{Wertz2012,Kammann2012}. While Josephson oscillations~\cite%
{Lagoudakis2010,Abbarchi2013} and other spatially dependent oscillations~%
\cite{DeGiorgi2014} were reported recently, the study of exciton-polariton
Rabi oscillations has been typically kept separate from the study of spatial
dynamics. This is likely due to the fact that Rabi oscillations are
short-lived, surviving only a limited number of cycles due to the short
polariton lifetime (a few tens of picoseconds in state-of-the art samples).
Nevertheless propagating polaritons have been progressing steadily toward
the realization of optical circuits, where their light effective mass and
strong nonlinear interactions have allowed several implementations of
optical switches~\cite{Amo2010,Adrados2011,DeGiorgi2012} and transistors~%
\cite{Gao2012,Ballarini2013}.

To overcome the limited duration of Rabi oscillation, one can consider the
amplification~\cite{Wertz2012} of polaritons by a non-resonant excitation.
This creates a reservoir of hot excitons, which can undergo stimulated
scattering into polariton states. The result is an effective incoherent
pumping or gain mechanism of polariton states, which can compensate
polariton decay~\cite{Demirchyan2014}. Using a Ginzburg-Landau type model~%
\cite{Keeling2008} we show that this results in sustained Rabi oscillations,
which brings new opportunities for their control, manipulation and
application.

Exciton-polaritons also have a rich spin dynamics~\cite{Shelykh2010},
allowed by their two-component spin degree of freedom. We show that the
propagation of polariton spin oscillations induced by Rabi oscillations \cite%
{Brunetti2006} in space can be influenced by applied magnetic fields, as
well as transverse electric-transverse magnetic (TE-TM) splitting of the
modes. We show that Rabi oscillations can be further controlled by applying
additional pulses to the system, which may enhance or suppress oscillations,
where the pulse phase becomes a control variable.

Finally, we consider oscillations between exciton-polariton states with
different momenta (i.e., different in-plane wavevectors), where propagating
domains in real space are distinguished by their phase. In analogy to
previous studies of domain wall propagation~\cite{Liew2008}, the domains act
as information carriers and logic gates can be realized from the combination
of domains at engineered points of space. However, unlike previous work, the
phase of the domains is a free continuous variable, opening an area of
analogue information processing in polaritonics.

\textit{Theoretical Model.---} To describe a coherent state of excitons and
cavity photons, we introduce the mean-field wavefunctions~\cite%
{Carusotto2004} of spin-polarized excitons, $\chi _{\sigma }$, and photons, $%
\phi _{\sigma }$. The index $\sigma =\pm $ accounts for the two possible
spin projections of (optically active) excitons and photons on the structure
growth axis. The evolution of the mean-fields is described by complex
Ginzburg-Landau equations~\cite{Wouters2007,Keeling2008}:
\begin{align}
i\hbar \frac{\partial \chi _{\sigma }}{\partial t}& =\left( E_{X}+\sigma
\Omega _{Z}+i\left( P_{\sigma }-\Gamma _{X}-\Gamma _{\mathrm{NL}}|\chi
_{\sigma }|^{2}\right) \right.   \notag \\
& \hspace{5mm}\left. +\alpha |\chi _{\sigma }|^{2}\right) \chi _{\sigma
}+\Omega \phi _{\sigma },  \label{eq:GPchi} \\
i\hbar \frac{\partial \phi _{\sigma }}{\partial t}& =\left( -\frac{\hbar ^{2}%
}{2m_{C}}\hat{\nabla}^{2}-i\Gamma _{C}\right) \phi _{\sigma }+\Omega \chi
_{\sigma }+F_{\sigma } .  \label{eq:GPphi}
\end{align}%
Here $E_{X}$ represents the exciton-photon detuning and we neglected the
dispersion of excitons, which is flat compared to the parabolic photon dispersion
given by the light photon effective mass $m_{C}$. In a magnetic field
excitons experience a Zeeman splitting~\cite{Shelykh2010}, given by $2\Omega
_{Z}$. The incoherent pumping of the system is described by the polarized~%
\cite{Kammann2012} pumping strength $P_{\sigma }$, which is saturated at
high densities due to non-linear losses~\cite{Keeling2008} characterized by $%
\Gamma _{\mathrm{NL}}$. We also allow for a coherent resonant pumping with
amplitude $F_{\sigma }$. $\Gamma _{X}$ and $\Gamma _{C}$ are the decay rates
of excitons and photons, respectively. Non-linear interactions between
excitons with parallel spins are introduced in Gross-Pitaevskii form~\cite%
{Carusotto2004} and described by the scattering strength $\alpha $. For simplicity, we
neglect the much weaker interactions between excitons with opposite
spins~\cite{Ciuti1998}. Finally, $\Omega $ is the Rabi coupling strength between the
excitons and photons.

\textit{Homogeneous Solutions.---} To gain some understanding of the states
supported by Eqs.~(\ref{eq:GPchi}) and~(\ref{eq:GPphi}), let us first
consider a spatially homogeneous incoherent pumping and no coherent pumping (%
$F_{\sigma }=0$). For simplicity, let us also first neglect the Zeeman
splitting and polariton-polariton interaction terms ($\Omega _{Z}=0$; $%
\alpha =0$). For pump powers exceeding a threshold $P\geq \Gamma _{C}+\Gamma
_{X}$, stationary homogeneous states exist:
\begin{equation}
\phi _{\sigma }=\frac{\Omega \chi _{\sigma }}{\hbar \mu +i\Gamma _{C}},%
\hspace{2mm}\chi _{\sigma }=\sqrt{\frac{P_{\sigma }-\Gamma _{C}-\Gamma _{X}}{%
\Gamma _{\mathrm{NL}}}}e^{-i\mu t+i\theta _{\sigma }}
\label{eq:HomogeneousSteady}
\end{equation}%
where $\hbar \mu =\sqrt{\Omega ^{2}-\Gamma _{C}^{2}}$. Since there are no
terms coupling $\sigma ^{+}$ and $\sigma ^{-}$ polarized states in Eqs.~(\ref%
{eq:GPchi}) and (\ref{eq:GPphi}), we effectively have two scalar problems.
This would not be the case in the presence of a polarization splitting~\cite%
{Supplemental}. Note that the incoherent pumping does not fix the phase of the solutions,
given by $\theta _{\sigma }$, which would be set by initial excitation conditions. The
stability of the stationary solution can be
checked by considering the spectrum of elementary excitations~\cite%
{Carusotto2004}. While the solution~(\ref{eq:HomogeneousSteady}) is stable,
it is not the only possibility.

\begin{figure}[t]
\centering
\includegraphics[width=8.5cm]{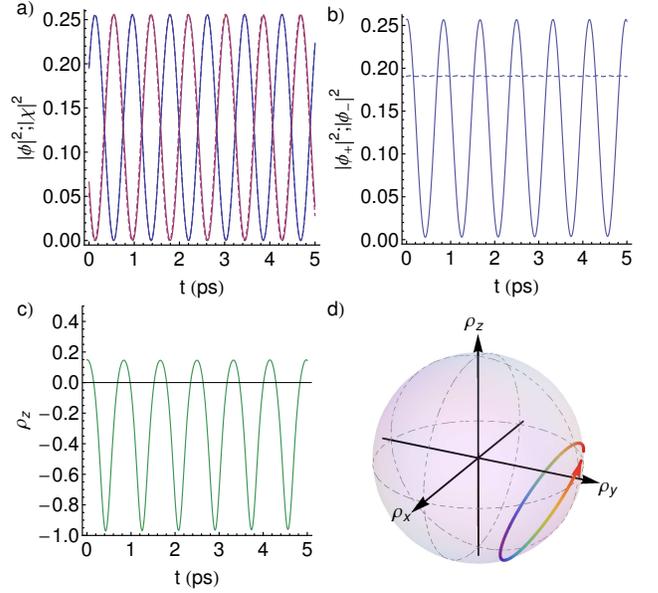}
\caption{(color online) a) Homogeneous oscillating solutions for the photon
and exciton intensity, $|\protect\phi(t)|^2$ and $|\protect\chi(t)|^2$,
respectively in the case of a scalar condensate (circularly polarized
excitation). Solid curves show numerically obtained results via the
application of a pulse to the system. Superimposed dashed curves show the
approximate analytical result. b) Under a linearly polarized pump, a
circularly polarized pulse generates exciton-photon Rabi oscillations in one
circularly polarized component while the other component maintains a fixed
intensity. This behaviour persists in the presence of Zeeman splitting. c)
Evolution of the (photonic) circular polarization degree $\protect\rho_z$.
d) Pseudospin evolution in the Poincar\'e sphere. Parameters: $\Omega=2.5$%
meV, $\hbar/\Gamma_C=10$ps, $\hbar/\Gamma_X=100$ps, $\Omega_Z=0.05$meV, $%
P=0.13$meV.}
\label{fig:Homogeneous}
\end{figure}

We may also consider oscillating solutions of the form $\chi _{\sigma
}(t)=\chi _{\sigma ,1}\sin (\omega t)$ and $\phi _{\sigma }(t)=i\phi
_{\sigma ,1}\sin (\omega t)+i\phi _{\sigma ,2}\cos (\omega t)$, where $\chi
_{\sigma ,1}$, $\phi _{\sigma ,1}$ and $\phi _{\sigma ,2}$ are taken to be
constants. Substituting into Eqs.~(\ref{eq:GPchi}) and~(\ref{eq:GPphi}), and
collecting terms oscillating as $\cos (\omega t)$ and $\sin (\omega t)$, we
obtain the approximate solution:
\begin{align}
\phi _{\sigma ,1}& =-\frac{\Omega \Gamma _{C}}{(\hbar \omega )^{2}+\Gamma
_{C}^{2}}\chi _{\sigma ,1}, & \phi _{\sigma ,2}& =\frac{\hbar \omega }{%
\Omega }\chi _{\sigma ,1},  \notag \\
|\chi _{\sigma ,1}|^{2}& =\frac{2}{3}\frac{\left( P-\Gamma _{C}-\Gamma
_{X}\right) }{\Gamma _{\mathrm{NL}}}, & \hbar \omega & =\pm \sqrt{\Omega
^{2}-\Gamma _{C}^{2}}  \label{eq:HomogeneousOscillating}
\end{align}%
Here we neglected fast oscillating terms proportional to $\sin (3\omega t)$,
appearing in the expansion of the term $\sin ^{3}(\omega t)=\left( 3\sin
(\omega t)-\sin (3\omega t)\right) $. A comparison with the direct numerical
solution of Eqs.~(\ref{eq:GPchi}) and~(\ref{eq:GPphi}) is shown in Fig.~\ref%
{fig:Homogeneous}. To place the system in an oscillating state a pulse $F_{\sigma }(t)$ was applied to the photon evolution according to Eq.~(%
\ref{eq:GPphi}). The Rabi oscillating solution is metastable. The phase difference
between the exciton and photon component is $\pi/2$ at any moment of time for this
solution. Fluctuations of the phase difference away from this value will lead to decay of
oscillations. However, the amplification provided by the continuous incoherent pump $P$
ensures that the oscillations are sustained long after the pulse has passed.

Given that the two spin polarized components are decoupled, it is possible
to prepare them in different states. Pairing a linearly polarized incoherent
pump ($P_+=P_-$) with a circularly polarized pulse $F_+(t)$ excites Rabi
oscillations in the $\sigma^+$ polarization, while the $\sigma^-$
polarization achieves the stationary state given by Eq.~(\ref%
{eq:HomogeneousSteady}). Such a situation is shown in Fig.~\ref%
{fig:Homogeneous}b, and is qualitatively unchanged even in the presence of a
magnetic field ($\Omega_Z\neq0$). In this way, the presence of Rabi
oscillations also manifests in oscillations in the circular polarization
degree, $\rho=\frac{|\phi_+|^2-|\phi_-|^2}{|\phi_+|^2+|\phi_-|^2}$ (see Fig.~%
\ref{fig:Homogeneous}), and a circular rotation of the pseudospin vector~%
\cite{Shelykh2010} on the Poincar\'e sphere (Fig.~\ref{fig:Homogeneous}d).
In contrast to common expectations, this pseudospin rotation is not related
to any polarization splitting.

\begin{figure}[t]
\centering
\includegraphics[width=8.5cm]{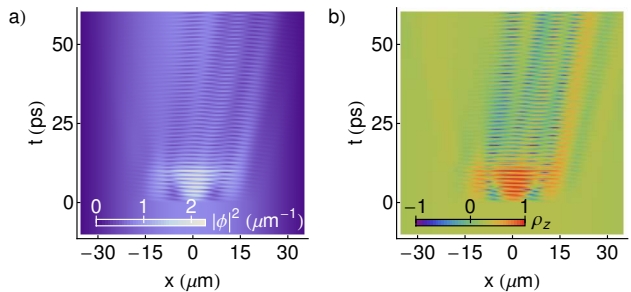}
\caption{(color online) a) Spatial dynamics of a polariton condensate
excited by a linearly polarized incoherent pump. A circularly polarized
pulse with non-zero in-plane wavevector induces propagating Rabi
oscillations, beginning at $t=0$ps. b) Associated dynamics of the circular
polarization degree. Parameters were the same as in Fig.~\protect\ref%
{fig:Homogeneous} with $m_C=7.5\times10^{-5}m_e$ and $k_F=0.5\protect\mu$m$%
^{-1}$.}
\label{fig:SpatialDynamics1}
\end{figure}

\begin{figure}[t]
\centering
\includegraphics[width=8.5cm]{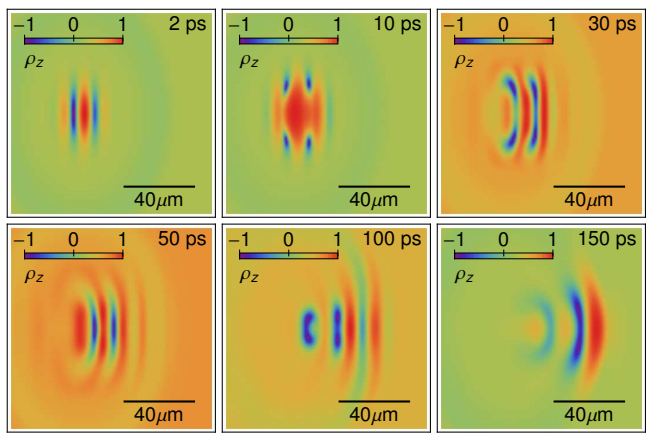}
\caption{(color online) Snapshots of the spatial distribution of the
circular polarization degree at selected times, showing the propagation of
Rabi oscillations corresponding to Fig.~\protect\ref{fig:SpatialDynamics1}.}
\label{fig:SpatialSnapshots}
\end{figure}

\textit{Spatial Propagation of Rabi Oscillations.---} If the applied
coherent pulse is not homogeneous in space, but rather localized, then one
can consider the resulting propagation of the induced spin-Rabi
oscillations. Figure~\ref{fig:SpatialDynamics1} shows results from the
numerical solution of Eqs.~(\ref{eq:GPchi}) and~(\ref{eq:GPphi}) using a
broad Gaussian shaped incoherent cw excitation and a focused Gaussian shaped
pulse. Fast oscillations can again be observed in the total polariton
intensity ($|\phi _{+}|^{2}+|\phi _{-}|^{2}$) and circular polarization
degree. However, the introduction of a non-zero in-plane wavevector of the
pulse (i.e., modulation of $F_{+}$ by $\exp (ik_{F}x)$) induces the spatial
propagation of oscillations, which continue toward the edge of the
incoherent pump even after the pulse has passed. This spreading of an
oscillating non-uniform spin polarization is further illustrated by several
snapshots of the spatial distribution of the circular polarization degree
shown in Fig.~\ref{fig:SpatialSnapshots}.
\begin{figure}[t]
\centering
\includegraphics[width=8.5cm]{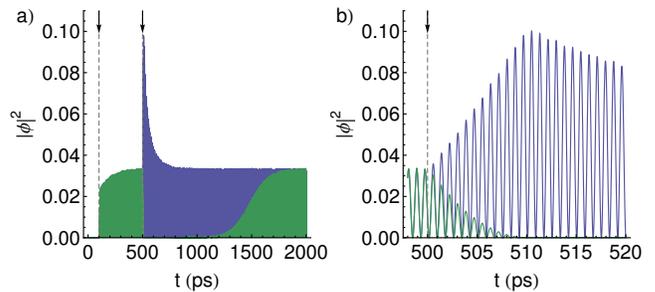}
\caption{(color online) Effect of two pulses on the scalar homogeneous
system. An initial pulse at $t=100$ps (marked by an arrow) induces Rabi
oscillations in the system. An additional pulse is applied at $t=500$ps
(marked by another arrow), which is either in phase (blue) or out-of phase
(green) with the Rabi oscillations. The in-phase pulse causes a temporary
amplification of oscillations while the out-of-phase pulse suppresses the
oscillations, which return after a delay. Parameters were the same as in
Fig.~\protect\ref{fig:Homogeneous}, but with $P=0.08$meV.}
\label{fig:HomogeneousPulse}
\end{figure}

\textit{Two-pulse Excitation.---} Let us now return to considering the
homogeneously excited scalar system (considering separately a particular
spin component). Figure~\ref{fig:HomogeneousPulse} shows the time evolution
of the photon intensity $|\phi |^{2}$ when subjected to a pair of pulses
arriving at times indicated by the arrows and vertical dashed lines. The
first pulse induces Rabi oscillations in the system, which reach a fixed
maximum amplitude (as seen before in Fig.~\ref{fig:Homogeneous}). The second
pulse can either amplify or suppress the oscillations, depending on whether
it is in-phase or out-of-phase with the oscillations. In the former case,
the amplification decays quickly, on the order of the polariton lifetime. In
the case of an out-of-phase second pulse, the Rabi oscillations in the
system can be suppressed for an extended period, with careful tuning of the
pulse amplitude.

\begin{figure}[t]
\centering
\includegraphics[width=8.5cm]{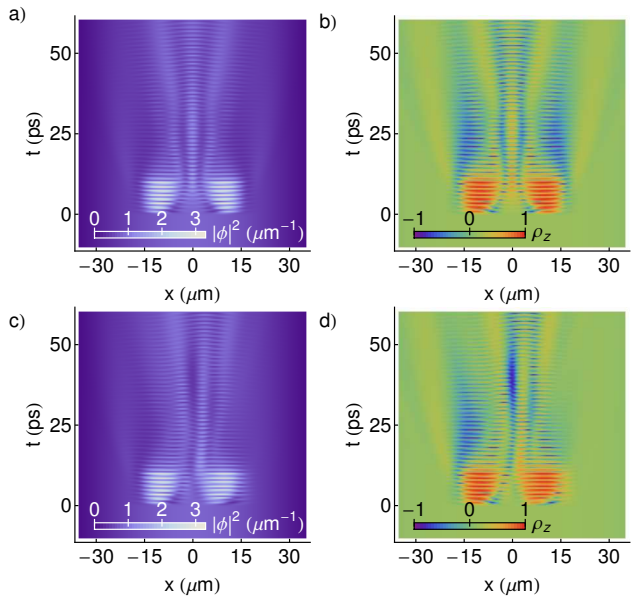}
\caption{(color online) Interference of Rabi oscillations generated by two
spatially separated pulses. Each pulse arrives at the same time, within the
area of a background Gaussian shaped incoherent continuous wave pump. a) and
b) Evolution of the total intensity and circular polarization degree,
respectively, for in-phase pulses. c) and d) the same for out-of-phase
pulses. Parameters were the same as in Fig.~\protect\ref%
{fig:SpatialDynamics1}. The two pulses have equal and opposite wavevectors, $%
k_F=\pm0.5\protect\mu$m$^{-1}$, and are shifted by $x=\mp 10\protect\mu$m,
respectively, from the center of the incoherent pump. }
\label{fig:TwoPulses}
\end{figure}
\textit{Interactions between propagating Rabi domains.---} In the case of
spatially separated pulses, with Gaussian spatial profiles, it is possible
to observe collisions between propagating Rabi domains. In Fig.~\ref%
{fig:TwoPulses}, two Rabi oscillating domains are generated by a pair of
pulses arriving at $t=0$ps. The pulses set the phase of their respective
domains, and the resulting interference pattern of the domains differs
depending on whether the pulses arrive in-phase (Fig.~\ref{fig:TwoPulses}a,b) or
out-of-phase (Fig.~\ref{fig:TwoPulses}c,d).

This phase sensitivity of interfering polaritons has previously been
appreciated as an ingredient for polaritonic information processing~\cite%
{Shelykh2009}. The spreading of Rabi oscillating domains is also reminiscent
of spin polarized domains, which can be used to realize binary logic gates~%
\cite{Liew2008}. However, a limitation of the propagation shown in Fig.~\ref%
{fig:SpatialDynamics1} is that it occurs only in one direction - the one set
by the wavevector of the applied pulse. For the construction of cascadable
logic gates, one typically needs to have signals capable of travelling in a
variety of directions in the microcavity plane.

\begin{figure}[b]
\centering
\includegraphics[width=8.5cm]{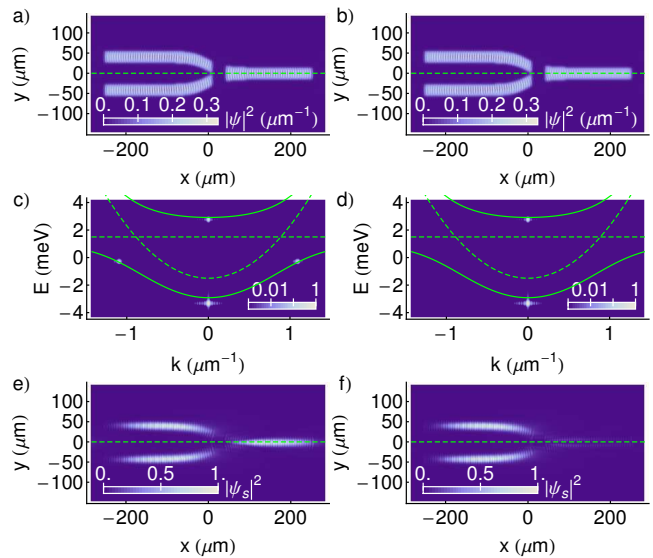}
\caption{(color online) Interference of parametric oscillations in polariton
channels formed by spatial patterning of the pumping field $F(x)$. a) and b)
show the intensity distribution of polaritons $200$ps after the arrival of a
pair of input pulses applied to the left-hand channels with the same and
opposite phases, respectively. c) and d) show the polariton dispersions
along the horizontal dashed lines in (a) and (b), respectively. Dashed
curves represent the bare exciton (flat) and photon dispersion, while solid
curves show the lower and upper polariton branches. e) and f) show the
polariton signal intensity, obtained by filtering around the wavevector $%
k_s\approx1.05\protect\mu$m$^{-1}$) in (c) and (d), respectively. }
\label{fig:Logic}
\end{figure}

\textit{Parametric Oscillations and Analogue Logic in Real Space.---} To
allow for oscillating domains that propagate in all directions, we make use
of the potential parametric scattering~\cite{Savvidis2000} between polariton
modes, allowed by the nonlinear $\alpha $-dependent term in Eq.~\ref%
{eq:GPchi}. Considering the simultaneous coherent excitation of the bottom
of the lower and upper polariton branches, one can expect scattering to
non-zero wavevector states on the lower polariton branch~\cite{Ciuti2004},
as demonstrated in Fig.~\ref{fig:Logic}c (this requires a positive
exciton-photon detuning; $E_{X}=3$meV).

By the spatial patterning of the pumping field $F(x)$, one can confine
polaritons along channels and we consider a ``Y''-shaped channel in Fig.~\ref%
{fig:Logic}. Applied pulses localized in the left-hand channels trigger the
parametric scattering and also set the phase of the resulting signal states.
Figure~\ref{fig:Logic}a shows the polariton intensity $200$ps after the
pulses arrive, which are chosen to have the same phase (for simplicity, we
consider only a single spin component here). A weak spatial modulation of
the polariton density is associated with the scattering in reciprocal space
shown in Fig.~\ref{fig:Logic}c. Filtering of the polariton field around the
signal wavevector ($k_s\approx1.05\mu$m$^{-1}$) clearly shows that the
signal has propagated into the right-hand output channel.

In contrast, when pulses excite signals in the channels with opposite phase,
they interfere destructively at the point where channels join, suppressing
the propagation.

A further advantage of this scheme is that the signal wavevector can be
tuned near the point of maximum group velocity of the lower polariton
dispersion. In principle, this allows repetition rates of the order of tens
of gigahertz, which could be further improved by reducing the dimensions of
the channel pattern. On the other hand, oscillating parametric polariton solitons~\cite{Egorov2013,Egorov2014} with very different spatial profiles and frequencies can be obtained operating near a flatter exciton-like region of the dispersion

\textit{Conclusion.---} We considered the generation of exciton-polariton
Rabi oscillating domains in semiconductor microcavities subjected to
coherent pulses. A continuous wave incoherent pump compensates the polariton
lifetime, giving rise to sustained oscillations. The spin polarization of
oscillations can be selected via the pulse polarization, which also allows
the generation of terahertz frequency oscillations in the polariton spin
degree of freedom. The oscillations remain in the presence of magnetic
fields or TE-TM splitting, and can be further controlled by the application
of additional pulses.

An important property of the Rabi domains is that their phase can be varied,
which gives a continuous variable for encoding information. By making use of
interbranch polariton-polariton scattering, the propagation of oscillating
domains can be controlled along channels by patterning the incident optical
field. A logical phase-dependent behaviour is observed from the interference
when domains collide. This opens a route for analogue architectures in
polaritonic devices.

\textit{Acknowledgements. } We are indebted to D. Sanvitto, F. Laussy, A. Alodjants, I.
A. Shelykh, and K. Kristinsson for fruitful discussions. YGR and AVK acknowledge
support from the EU FP7 IRSES Project POLAPHEN. AK also acknowledges support from the EPSRC Established Career Fellowship.

\clearpage

\begin{centering}
{\bf SUPPLEMENTARY INFORMATION}\\
\end{centering}
\setcounter{equation}{0} \setcounter{figure}{0} \renewcommand{%
\theequation}{S\arabic{equation}} \renewcommand{\thefigure}{S\arabic{figure}}

\textit{Influence of TE-TM splitting on propagating spin-Rabi
oscillations.---} Many semiconductor microcavities exhibit an additional
k-dependent polarization splitting, mainly due to the different energies of
transverse electric (TE) and transverse magnetic (TM) photonic cavity modes.
This splitting is well-known to influence the spin dynamics and resulting
spatial patterns formed by propagating polaritons (see e.g.,~\cite%
{Shelykh2010,Kammann2012} and references within).

Theoretically, the TE-TM splitting can be accounted for by adding a
wavevector dependent term that couples the two spin components to the right
hand side of Eq.~(\ref{eq:GPphi}):
\begin{equation}
i\hbar\frac{\partial\phi_\sigma}{\partial t}=\ldots+\frac{\Delta_\mathrm{LT}%
}{k^2_\mathrm{LT}}\left(i\frac{\partial}{\partial x}+\sigma\frac{\partial}{%
\partial y}\right)\phi_{-\sigma}
\end{equation}
where $\Delta_\mathrm{LT}$ determines the strength of the TE-TM splitting at
the in-plane wavevector $k_\mathrm{LT}$.

For typical parameters, we obtain the result shown in Fig.~\ref%
{fig:SpatialSnapshotsTETM}. The basic phenomenology of propagating spin-Rabi
oscillations remains, however, the TE-TM splitting breaks the mirror
symmetry of the system about the horizontal axis. This gives rise to
assymmetric patterns in the distribution of the circular polarization
degree.
\begin{figure}[h]
\centering
\includegraphics[width=8.5cm]{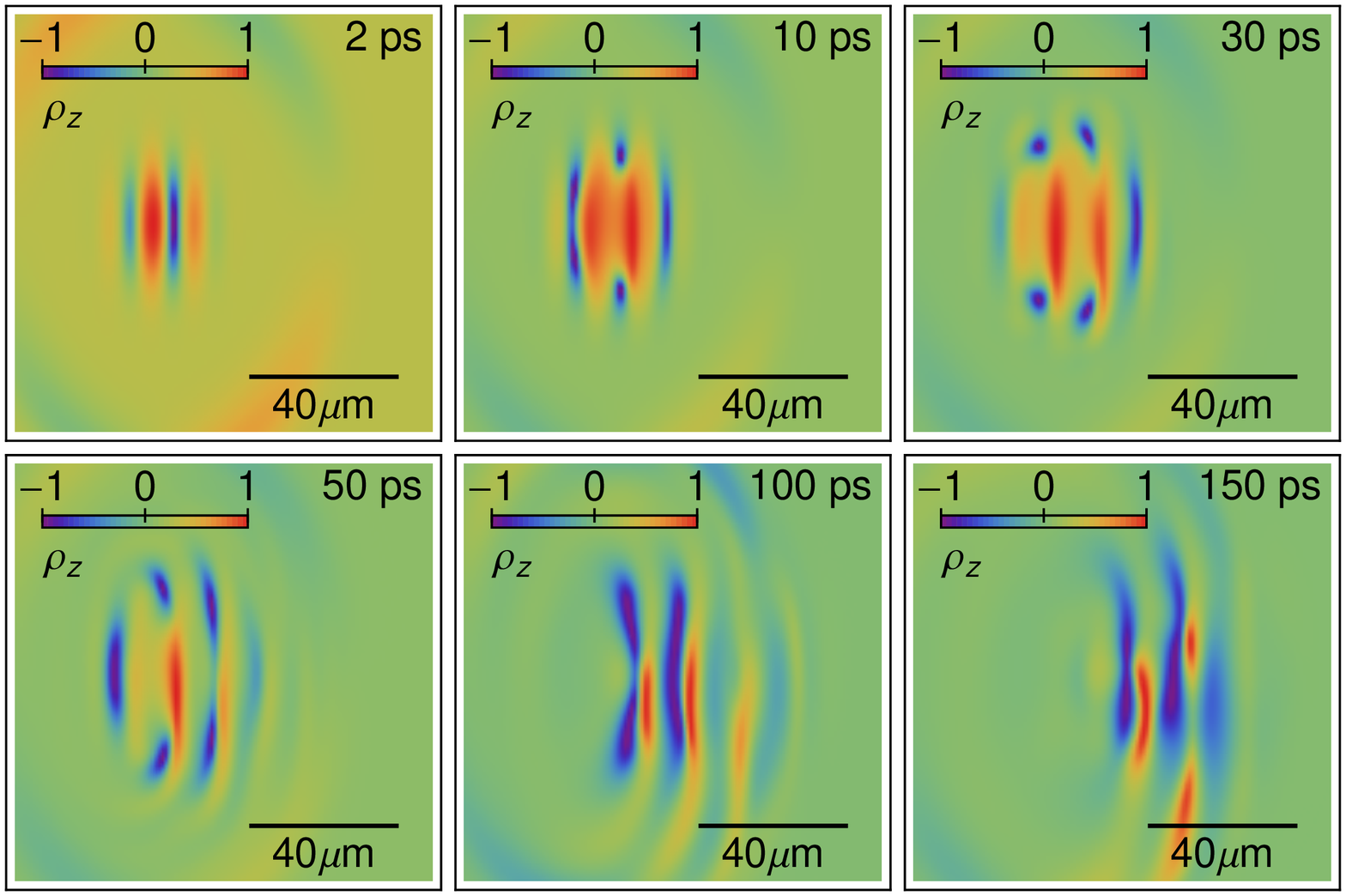}
\caption{(color online) Influence of TE-TM splitting, giving rise to an
asymmetric spatial distribution of Rabi oscillations. Images show the same
as in Fig.~\protect\ref{fig:SpatialSnapshots}, accounting for TE-TM
splitting. Parameters: $\Delta_\mathrm{LT}=0.1$meV, $k_\mathrm{LT}=1\protect%
\mu$m$^{-1}$, $\Omega_Z=0$meV.}
\label{fig:SpatialSnapshotsTETM}
\end{figure}


\begin{thebibliography}{99}
\bibitem{Zoller1995} J I Cirac \& P Zoller, Phys. Rev. Lett., \textbf{74},
4091 (1995).

\bibitem{Leibrandt2009} D R Leibrandt, J Labaziewicz, R J Clark, I L Chuang,
R J Epstein, C Ospelkaus, J H Wesenberg, J J Bollinger, D Leibfried, D J
Wineland, D Stick, J Sterk, C Monroe, C S Pai, Y Low, R Frahm, R E Slusher,
Quantum Inf. Comput., \textbf{9}, 901 (2009).

\bibitem{Kaluzny1983}
Y Kaluzny, P Goy, M Gross, J M Raimond, \& S Haroche, Phys. Rev. Lett.,
\textbf{51}, 1175 (1983).

\bibitem{Martinis2002}
J M Martinis, S Nam, J Aumentado, \& C Urbina, Phys. Rev. Lett., \textbf{89}%
, 117901 (2002).

\bibitem{Yu2002}
Y Yu, S Y Han, X Chu, S I Chu, \& Z Wang, Science, \textbf{296}, 889 (2002).

\bibitem{Petta2005}
J R Petta, A C Johnson, J M Taylor, E A Laird, A Yacoby, M D Lukin, C M
Marcus, M P Hanson, \& A C Gossard, Science, \textbf{309}, 2180 (2005).

\bibitem{Koppens2006}
F H L Koppens, C Buizert, K J Tielrooij, I T Vink, K C Nowack, T Meunier, L
P Kouwenhoven, L M K Vandersypen, Nature, \textbf{442}, 766 (2006).

\bibitem{Jelezko2004}
F Jelezko, T Gaebel, I Popa, M Domhan, A Gruber, \& J Wrachtrup, Phys. Rev.
Lett., \textbf{93}, 130501 (2004).

\bibitem{Yang2012}
J Yang, Y Wang, Z Wang, X Rong, C K Duan, J H Su, \& J Du, Phys. Rev. Lett,
\textbf{108}, 230501 (2012).

\bibitem{Schulzgen1998}
A Sch\"ulzgen, R Binder, M E Donovan, M Lindberg, K Wundke, H M Gibbs, G
Khitrova, \& N Peyghambarian, Phys. Rev. Lett., \textbf{82}, 2346 (1999).

\bibitem{Norris1994}
T B Norris, J K Rhee, C Y Sung, Y Arakawa, M Nishioka, \& C Weisbuch, Phys.
Rev. B, \textbf{50}, 14663 (1994).

\bibitem{Brunetti2006}
A Brunetti, M Vladimirova, D Scalbert, M Nawrocki, A V Kavokin, I A Shelykh, \& J Bloch, Phys. Rev. B, {\bf 74}, 241101 (2006).

\bibitem{High2013} A A High, A T Hammack, J R Leonard, Sen Yang, L V Butov,
T Ostatnicky, M Vladimirova, A V Kavokin, T C H Liew, K L Campman, \& A C
Gossard, Phys. Rev. Lett., \textbf{110}, 246403 (2013).

\bibitem{Wertz2012} E Wertz, A Amo, D D Solnyshkov, L Ferrier, T C H Liew, D
Sanvitto, P Senellart, I Sagnes, A Lema\^itre, A V Kavokin, G Malpuech, \& J
Bloch, Phys. Rev. Lett., \textbf{109}, 216404 (2012).

\bibitem{Kammann2012} E Kammann, T C H Liew, H Ohadi, P Cilibrizzi, P
Tsotsis, Z Hatzopoulos, P G Savvidis, A V Kavokin, \& P G Lagoudakis, Phys.
Rev. Lett., \textbf{109}, 036404 (2012).

\bibitem{Lagoudakis2010} K G Lagoudakis, B Pietka, M Wouters, R Andre, \& B
Deveaud-Pledran, Phys. Rev. Lett., \textbf{105}, 120403 (2010).

\bibitem{Abbarchi2013} M Abbarchi, A Amo, V G Sala, D D Solnyshkov, H
Flayac, L Ferrier, I Sagnes, E Galopin A Lema\^itre, G Malpuech, \& J Bloch,
Nature Phys., \textbf{9}, 275 (2013).

\bibitem{DeGiorgi2014} M De Giorgi, D Ballarini, P Cazzato, G Deligeorgis, S
I Tsintzos, Z Hatzopoulos, P G Savvidis, G Gigli, F P Laussy, \& D Sanvitto,
Phys. Rev. Lett., \textbf{112}, 113602 (2014).

\bibitem{Amo2010} A Amo, T C H Liew, C Adrados, R Houdr\'e, E Giacobino, A V
Kavokin, \& A Bramati, Nature Photon., \textbf{4}, 361 (2010).

\bibitem{Adrados2011} C Adrados, T C H Liew, A Amo, M D Mart\'in, D Sanvitto,
C Ant\'on, E Giacobino, A Kavokin, A Bramati, \& L Vi\~na, Phys. Rev. Lett.,
\textbf{107}, 146402 (2011).

\bibitem{DeGiorgi2012} M De Giorgi, D Ballarini, E Cancellieri, F M
Marchetti, M H Szymanska, C Tejedor, R Cingolani, E Giacobino, A Bramati, G
Gigli, \& D Sanvitto, Phys. Rev. Lett., \textbf{109}, 266407 (2012)

\bibitem{Gao2012} T Gao, P S Eldridge, T C H Liew, S I Tsintzos, G
Stavrinidis, G Deligeorgis, Z Hatzopoulos, \& P G Savvidis, Phys. Rev. B,
\textbf{85}, 235102 (2012).

\bibitem{Ballarini2013} D Ballarini, M De Giorgi, E Cancellieri, R Houdre, E
Giacobino, R Cingolani, A Bramati, G Gigli, \& D Sanvitto, Nature Comm.,
\textbf{4}, 1778 (2013).

\bibitem{Demirchyan2014} S S Demirchyan, I Y Chestnov, A P Alodjants, M M
Glazov, \& A V Kavokin, Phys. Rev. Lett., \textbf{112}, 196403 (2014).

\bibitem{Keeling2008} J Keeling \& N G Berloff, Phys. Rev. Lett., \textbf{100%
}, 250401 (2008).

\bibitem{Shelykh2010} I A Shelykh, Y G Rubo, A V Kavokin, T C H Liew, \& G
Malpuech, Semicond. Sci. Technol., \textbf{25}, 013001 (2010).

\bibitem{Liew2008} T C H Liew, A V Kavokin, \& I A Shelykh, Phys. Rev.
Lett., \textbf{101}, 016402 (2008).

\bibitem{Carusotto2004} I Carusotto \& C Ciuti, Phys. Rev. Lett., \textbf{93}%
, 166401 (2004).

\bibitem{Wouters2007} M Wouters \& I Carusotto, Phys. Rev. Lett., \textbf{99}%
, 140402 (2007).

\bibitem{Ciuti1998} C Ciuti, V Savona, C Piermarocchi, A Quattropani, \& P
Schwendimann, Phys. Rev. B, \textbf{58}, 7926 (1998).

\bibitem{Supplemental} See the Supplemental Material for a comparison to the
case including TE-TM splitting.

\bibitem{Shelykh2009} I A Shelykh, G Pavlovic, D D Solnyshkov, \& G
Malpuech, Phys. Rev. Lett., \textbf{102}, 046407 (2009).

\bibitem{Savvidis2000} P G Savvidis, J J Baumberg, R M Stevenson, M S
Skolnick, D M Whittaker, \& J S Roberts, Phys. Rev. Lett., \textbf{84}, 1547
(2000).

\bibitem{Ciuti2004} C Ciuti, Phys. Rev. B, \textbf{69}, 245304 (2004).

\bibitem{Egorov2013}
O A Egorov \& F Lederer, Phys. Rev. B, {\bf 87}, 115315 (2013).

\bibitem{Egorov2014}
O A Egorov \& F Lederer, Optics Lett., {\bf 39}, 4029 (2014).

\end{thebibliography}
\end{document}